\documentclass[10pt]{article}
\usepackage[utf8]{inputenc}
\usepackage[left=2cm, right=2cm, top=2cm]{geometry}
\usepackage{parskip} 
\usepackage{tikz} 
\usepackage{amsmath}
\usepackage{bm}
\usepackage{hyperref}
\usepackage{graphicx}
\usepackage{epstopdf}
\usepackage{float}
\usepackage[round,numbers,comma,sort]{natbib}

\setcitestyle{square}

\begin{document}

\title{A Model of the Optimal Selection of Crypto Assets\thanks{PRELIMINARY DRAFT.}}
\author{Silvia Bartolucci\thanks{The Centre for Global Finance and Technology at the Imperial College Business School and UCL Centre for Blockchain Technologies. Correspondence to: \texttt{s.bartolucci@imperial.ac.uk}} \and Andrei Kirilenko\thanks{The Centre for Global Finance and Technology at the Imperial College Business School, Cambridge Judge Business School and CEPR. Email: \texttt{a.kirilenko@imperial.ac.uk}}}

\date{}
\maketitle
\begin{abstract}
\noindent 
We propose a modelling framework for the optimal selection of crypto assets. Crypto assets differ by two essential features: {\em security} (technological) and {\em stability} (governance). Investors make choices over crypto assets similarly to how they make choices by using a recommender app: the app presents each investor with a pair of crypto assets with certain security-stability characteristics to be compared. Each investor submits its preference for adopting one of the two assets to the app. The app, in turn, provides a recommendation on whether the proposed adoption is sensible given the assets' essential features, information about the adoption choices of all other investors, and expected future economic benefits of adoption. Investors continue making their adoption choices over all pairs of crypto assets until their expected future economic benefits can no longer be improved upon. This constitutes an optimal selection decision. We simulate optimal selection decisions considering the behaviour of different types of investors, driven by their attitudes towards assets' features. We find a variety of possible emergent outcomes for the investments in the crypto-ecosystem and the future adoption of the crypto assets.
\\
\vspace{0in}
\vspace{0in}\\

\bigskip
\end{abstract}
\setcounter{page}{0}
\thispagestyle{empty}

\section{Introduction}

A crypto asset is an intangible digital asset whose issuance, sale or transfer are secured by cryptographic technology and shared electronically via a distributed ledger (blockchain). The blockchain, in turn, is a database of issuance and transaction records (ledger), copies of which are stored on multiple computing devices (nodes) that form a distributed computer network. Each crypto asset has its own blockchain supported by its own network of nodes that provide processing power and memory capacity. All nodes of a given blockchain store copies of the entire history of the blockchain. Some nodes also support a blockchain by selecting, validating, and adding/chaining new blocks of records to the ledger in accordance with a pre-specified (consensus) algorithm \cite{anto, tascageneral,astegeneral}. 

Starting with the issuance of Bitcoin on the bitcoin blockchain in 2008 \cite{satoshi}, thousands of crypto assets have been issued over the last decade. The crypto asset ecosystem currently encompasses assets with very diverse underlying technological features (specific cryptographic technologies and electronic sharing protocols), as well as varying governance solutions (private vs. open access to the ledger and a variety of consensus algorithms).

Under the current standards, a crypto asset does not meet the definition of either cash or a financial instrument because it does not represent a claim or contractual relationship that results in a monetary or financial liability on any identifiable entity \cite{aasb}. A crypto asset, however, is an intangible digital asset as it is without physical substance (intangible), but is digitally-identifiable (digital) and held in expectation of future economic benefits (asset). As such, crypto assets have been sold by issuers ranging from genuine to outright fraudulent, bought by scores of investors with different degree of sophistication, and, consequently, attracted the scrutiny of regulators worldwide. While the majority of crypto assets will become worthless, some could end up being adopted widely enough to ensure their survival. Furthermore, a very small number of crypto assets could — despite likely booms and busts in their prices — become preferred assets of the future used by large and small investors alike to store and transfer wealth in a cryptographically protected, intangible, digitally-native form.

How are investors going to be making optimal selection decisions over many available crypto assets? Which features of these intangible digital assets would drive investment choices? Which types of assets (that do not represent on-blockchain liabilities) will survive and which will go extinct?

In this paper, we propose a modelling framework for the selection among existing and future crypto assets. In our framework, crypto assets differ by two essential features: {\em security} and {\em stability}. Security of a crypto asset represents the technological sophistication of the cryptographic and electronic communication technologies used to withstand cyber fraud, manipulation, abuse, and attack. Use of a more advanced encryption technology would render a crypto asset more secure relative to other crypto assets. Stability of a crypto asset represents its potentially faulty {\em governance} that may allow for a loss, misappropriation or dilution of its value. Use of credible legal, regulatory, and self-regulatory (e.g., consensus mechanism) attributes makes a crypto asset more stable. For example, stability of a crypto asset could be improved if it can be credibly represented as an off-ledger liability on an identifiable entity such as a central bank, foundation, company or special purpose vehicle among others.

In our framework, investors make choices over crypto assets similarly to how they make choices by using a recommender app: the app presents each investor with a pair of crypto assets with certain security-stability characteristics to be compared; each investor submits its preference for adopting one of the two assets to the app; the app, in turn, provides a recommendation on whether the proposed adoption is sensible given the assets' essential features, information about the adoption choices of all other investors, and expected future economic benefits of adoption. Investors continue making their adoption choices over all pairs of crypto assets until their expected future economic benefits can no longer be improved upon, which constitutes optimal selection decisions.

We simulate investors’ optimal selection decisions on a landscape where four types of crypto assets are of main interest  - high security/high stability, low security/high stability, high security/low stability, and low security/low stability. For expositional purposes, we characterise them as central bank digital currencies (CBDCs), stablecoins, cryptocurrencies, and crypto tokens respectively. 

A {\em central bank digital currency}  (CBDC) can be defined as either a digitally native form of fiat currency of a country or a balance held in a digital form in a reserve account at the country’s monetary authority. If issued by a credible monetary authority, it could be deemed very stable as it would represent an outright (off-ledger) liability of the monetary authority. It can also be made very secure by using a combination of advanced cryptographic technologies, encryption algorithms and cyber defense capabilities administered by the centralised monetary authority  \cite{CBDC-BIS,CDBoC, CBando,CBando2}. CBDCs raise complex issues ranging from technological and economic to fundamental, e.g., what should the role of a central bank be in an increasingly digital world?

{\em Stablecoins} are crypto assets whose values are pegged to baskets of fiat currencies or cash equivalents, existing financial instruments, physical assets such as commodities, as well as baskets of other crypto assets \cite{SCreport}. There exist three main stablecoins categories depending on the collateralisation method. Asset-collateralised stablecoins are backed by (off-ledger) assets, e.g. fiat in USD or EUR, and are the most centralised, as they rely on a central authority serving as custodian of the assets used to back the crypto asset. Crypto-collateralised assets are more decentralised: the collateralisation is done on-chain, i.e. locking digital assets on a distributed ledger platform using smart contracts. Non-collateralised  stablecoins are algorithmically-backed assets, where their increase or decrease of coin supply in the system is mathematically determined and automatically modified.
	 
{\em Cryptocurrencies} are decentralised crypto assets relying on cryptography to secure the transfer of value between peers in the network \cite{anto}. The pioneering cryptocurrency, Bitcoin \cite{satoshi}, appeared in 2008 and was followed soon after by a remarkable number of other coins, presenting very heterogeneous characteristics and improving on a subset of functionalities of Bitcoin. Notable examples are the Ethereum platform, where smart contracts functionalities have been introduced, or Monero and ZCash platforms with improved user's anonymisation techniques and with new implemented cryptographic primitives enhancing the capabilities of the systems \cite{monero,zcash}.
	
{\em Crypto tokens} are tradable crypto assets and utilities built on distributed ledger platforms. Crypto tokens can be considered utility tokens if they grant holders access to an existing current or future product or service built on an existing distributed ledger platform, such as Ethereum. In some instances, tokens may be considered future sales, investment and participation schemes created to fund projects and may present features typical of securities, i.e. claims on future cash flows. Indeed, tokens that do fall outside this classification also exists and they may offer the most diverse rights and functionalities for the users \cite{tokenclass}.

Our paper contributes to the emerging literature on crypto assets and blockchains. One strand of the literature studies the features of different crypto assets classes with the aim to categorise them and predict their future behaviour. For example, in \cite{tasca}, the authors present a universal taxonomy to navigate and distinguish different blockchains and distributed ledgers according to the building blocks and sub-component of the platform (e.g. cryptography, consensus mechanism, etc.). In \cite{cryptoclass}, instead, crypto tokens and cryptocurrencies are classified according to their market capitalisation: the power law distribution of the crypto tokens market capitalisation presents a larger exponent compared to cryptocurrencies. The growth of the two classes of currencies and their market capitalisation growth and distribution were described via an analogous of a population dynamics model. 

Another strand of the literature studies economic incentives embedded in blockchain platforms, with particular reference to Bitcoin. In \cite{powecon, miningfee, aste} the incentive mechanism of mining fees and the costs associated to mining activities are investigated. In \cite{sockin, btcecon, btcecon2} modelling and analysis of cryptocurrencies incentives is performed and optimal cryptocurrencies designs are proposed. In our model, we do not specifically discuss and include low-level technological features of the platform (e.g. consensus or encryption protocols) but we provide a macro-classification of the assets in terms of two main attributes, security and stability.
	
There are also agent-based models of a cryptocurrency market. For example, \cite{marketsim} presents a Bitcoin market with different types of traders (chartist and/or random), reproducing some key stylised facts of the market, such as Bitcoin price typical time series and volatility clustering. In order to assess, instead, the mechanisms of acceptance of new digital currencies, in \cite{Luther} the authors show -- by using simple models of standard currency adoption, in particular \cite{Dowd} -- how network effects and switching costs may hinder the adoption of a new currency even if all agents agree on its superiority compared to existing ones. 	

Studies using empirical data focus on understanding the dynamics of the cryptocurrency market using machine learning techniques \cite{baronevo, baromachine}, exploring the properties and inefficiencies of  the Bitcoin peer-to-peer network \cite{dimatteo} and the evolution and categorisation of the network of payments on the Bitcoin blockchain \cite{TascaNetwork}.  Empirical research also include cryptocurrencies price forecast and valuation \cite{baroprediction} and sentiment analysis \cite{sentimentAste, tweets} to investigate the interplay between sentiment extracted from news (e.g. tweets) and cryptocurrencies prices.

The paper is organised as follows. In Sec. \ref{sec:model}, we introduce the crypto assets classification framework, describing the decision-making process of investors for crypto assets adoption. In Sec. \ref{sec:result}, we show the simulations results for the case of homogeneous investors and heterogeneous investors. Finally, in Sec. \ref{sec:conclusions} we discuss possible extensions and future outlooks.
\begin{figure}[htb!]
\centering
\includegraphics[width=0.7\textwidth]{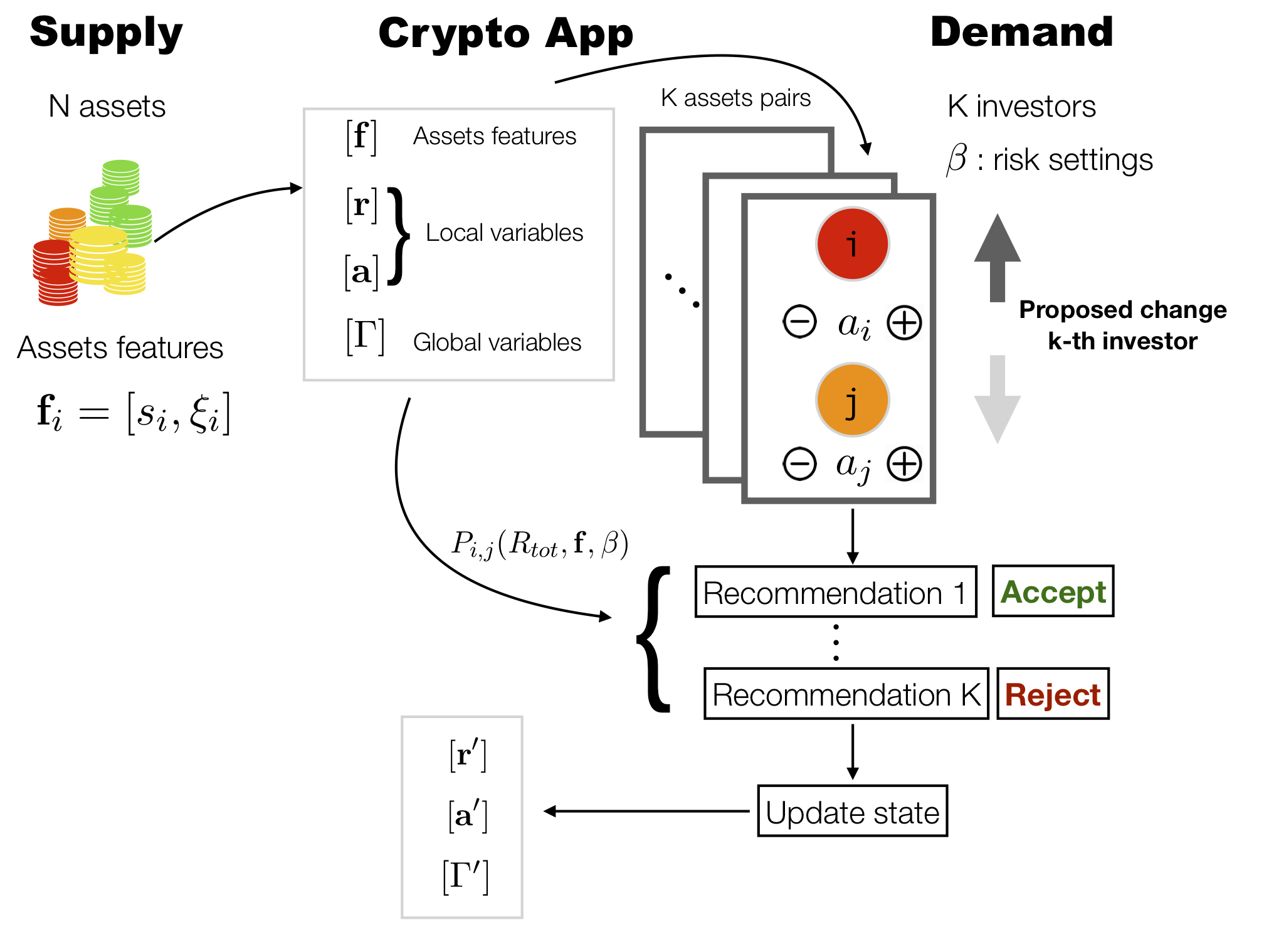}
\caption{Scheme of the main components of the optimal selection model.}
\label{fig:cryptoappscheme}
\end{figure}

\section{A Model of Crypto Assets} \label{sec:model}

In this section we introduce a model for the demand and supply of crypto assets available to investors optimised via a  ``crypto app''. The scheme highlighting the main components of the model, namely supply, demand and the features of the crypto app, is shown in Fig. \ref{fig:cryptoappscheme}.

\subsection{Supply of crypto assets} \label{sec:supply}

On the supply side, there are $N\gg1$ crypto assets made available to investors. We assume that crypto assets differ from each other by two essential features: {\em security} and {\em stability}. Security of a crypto asset represents its {\em technological} vulnerability to cyber fraud, manipulation, abuse, and attack. Use of a more advanced encryption technology would render a crypto asset more secure relative to other crypto assets at a point in time. In other words, security is a cross-sectional attribute of a crypto asset.\footnote{We assume that security improvements can be achieved by paying an exogenous fixed cost for a technological upgrade. Modelling this cost is a subject of future work.}

Stability of a crypto asset reflects its potentially faulty {\em governance}. Other things equal, greater reliance on (costly) regulated or self-regulated governance with a better defined legal or procedural recourse would render a crypto asset more stable in terms of the value that can be recovered. Thus, stability is a time series attribute of a crypto asset that reflects its ability to retain value across time for a given level of security. Improvements in governance can be achieved by adopting more credible legal, regulatory, and self-regulatory (e.g., consensus mechanism) attributes.\footnote{We assume that improvements in governance can be achieved and maintained by charging transaction fees. For the time being, we leave the mechanism for setting these transaction fees unspecified.} Stability of a crypto asset could be improved if it can be credibly represented as an off-ledger liability on an identifiable entity such as a central bank (e.g., in the case of a CBDC), a foundation, a limited liability legal entity or a special purpose vehicle among others. 

We consider security and stability as fixed exogenous attributes that take values in the intervals $s\in [0,1]$  and $\xi \in [0,1]$ respectively.\footnote{Without loss of generality, crypto assets with initial characteristics ($s$, $\xi$) can be obtained from the app store for free, because the code for their generation is available as ``open source.’’} Using this taxonomy, we can parametrise different classes of crypto assets. Assets can be defined by ranking them ranging from high stability and high security ones ($s, \xi\geq 0.5$) such as, for example, central bank digital currencies, to low stability and low security assets ($s, \xi < 0.5$) such as, for example, crypto tokens. Our classification of crypto assets is shown in table \ref{fig:tablec}. This rather simple framework allow us to categorise complex real-word assets in a quantitative and parsimonious way.

\begin{table}[H]
\centering
\includegraphics[width=0.5\textwidth]{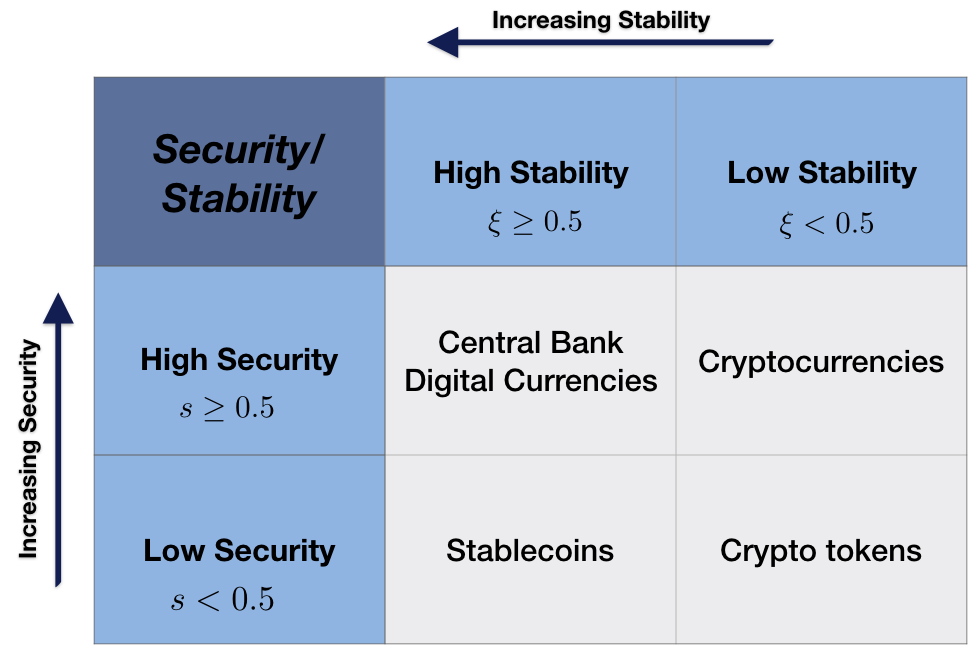}
\caption{Crypto assets classification according to their {\em security} ($s\in [0,1]$) and {\em stability} ($\xi \in [0,1]$) parameters.}
\label{fig:tablec}
\end{table}

\subsection{Demand for Crypto Assets} \label{sec:demand}
	Demand for crypto assets is specified below. At each point in time, $t$, a set of $K=\frac{N}{2}$ investors access the crypto app (described in detail in Sec. \ref{sec:app}) to search for and select crypto assets. The app provides each investor with a pair of crypto assets, $i$ and $j$ selected at random from the total pool. Each asset $i$ has known values of $r_i(t)$, representing its expected return\footnote{In general, the expected return of an asset $i$ at time $t$ is defined as
	$r_i(t) = E[x]=  \sum_{j=1}^n x_j(t) p_j(t)$, 
	and is a weighted sum of payoffs or returns $x_j(t)$ by the probability that the given return is achieved.  Note that in this paper, we will model only the dynamics at the level of the $r_i$'s.} and $a_i(t)$, representing its probability of adoption.

	At time $t$, the adoption process for crypto asset $i$ is a Bernoulli process: an investor can either adopt asset $i$ or not, giving rise to a probabilistic outcome that can be described by a binary variable $k=\{0,1\}$, which indicates that asset $i$ has not been ($k=0$) or has been ($k=1$) adopted (invested in), respectively. The probability that an investor adopts or does not adopt a given crypto asset $i$ can be described as
	\begin{equation}
	f(k, a_i (t)) = a_i^{k} (t) (1- a_i (t))^{1-k} \ ,
	\end{equation}
	where $a_i (t)$ is the probability of adoption for asset $i$ at time $t$. 

	The probability of adoption of asset $i$, $a_i (t), \ i=1,\dots,N$ is updated in time. The updating is executed by the crypto app, based on both local (investor-specific) and global parameters. At the local level, proposed binary choices of individual investors over all possible pairs of crypto assets are fed into the crypto app as follows: if an investor proposes to choose asset $j$ over asset $i$, the probability of adoption of asset $i$ goes down by a known quantity $\delta$ and the probability of adoption of asset $j$ goes up by $\delta$ (the probability adjustment is made symmetric to save on notation, without loss of generality). 

	Thus, for each pair of assets $(i,j)$, the proposed update for the probabilities of adoption reads:
	\begin{align}
	a_i(t+1)&= {\rm max}(a_i(t) -\delta,0)\ , \nonumber \\
	a_j(t+1)&= {\rm min}(a_j(t)+\delta,1) \ .
	\label{eq:aupdate}
	 \end{align}

	After collecting \emph{proposed} adjustments from all investors over all possible pairs of crypto assets, the app presents each investor with a recommendation (which may include investor-specific settings, see Sec. \ref{sec:settings}) to select $j$ over $i$ or not. Investors may choose the app’s recommendation at face value or may make their own choices depending on their trust in the app (the trust is not explicitly modelled in this paper). After investors made their choices, the app [sequentially] calculates the new vector of probabilities of adoption for each asset. Then, the app also updates expected returns based on the new probabilities of adoption and global parameters, as detailed below. 

\subsection{The Crypto App}  \label{sec:app}
Investors ``interact" with crypto assets over a digital platform, that we define as ``crypto app''.
The crypto app essentially serves as an optimising recommendation tool for investors. It stores information about available crypto assets, which are provided to its users wishing to make investment decisions over those assets. The app collects data about all users adoption preferences, which are then processed and used to provide investor-specific suggestions.
	It calculates aggregate information about the overall state of the ``market", and it provides optimal recommendations to agents in the space $[a, r]$, i.e. for assets probabilities of adoption and for their expected returns, given the values of adoption probabilities, as we will show in the following sections.
\subsubsection{Specifications}  \label{sec:spec}
As a standard procedure, upon signing up to the service, users are provided with specifications describing (i) the type of data collected and provided and (ii) how those data will be used to provide investment recommendations.

	The app stores information about the $N$ available assets together with their associated features $s_i,\xi_i \quad \forall i=1, \dots,N$, as well as their position in the adoption-expected return space at each point in time. The app also computes a global parameter, the total expected return, $R_{tot}(t)\in \Gamma(t)$, defined as 
	\begin{equation}
 	R_{tot}(t)= \sum_{i=1}^N a_i(t)r_i(t)\ , 
 	\label{eq:rtot}
 	\end{equation}
where $a_i(t)$ represents the $i$-th asset adoption at time $t$ and the quantity $r_i(t)$ is its expected return. This global parameter will be used by the app to evaluate whether an investor's choice to decrease (or increase) their propensity towards adopting a given asset will induce an overall decrease or increase in total expected return.
	
	Once an investor proposes a change in adoption, as described in Eq. \eqref{eq:aupdate}, the app calculates the total expected return $R_{tot}(t)$  if the proposed changes were adopted. The app also calculates the difference $\Delta R_{tot}(t) = R_{tot}(t-1)-R_{tot}(t)$ in total expected return, which is used to provide a recommendation to investors on whether or not they should proceed with the suggested choice. 
	
	We assume that the probability of accepting the changes proposed in Eq. \ref{eq:aupdate}, $P_{i,j}(\Delta R_{tot},{\bf f})$, directly depends on the difference between the total expected returns, $\Delta R_{tot}(t)$,  before and after modifying the adoptions $a_i(t)$, as well as on the intrinsic features of the crypto assets being compared, $\bf f $, i.e. their {\em stability} and {\em security} parameters. 
		
	We assume that the app uses the following functional form to calculate the probability of recommending the changes in adoption between assets $i,j$ and thus transitioning from the old states $a_i, a_j$ to the new proposed ones $\tilde{a}_i,  \tilde{a}_j$ (according to the update in Eq. \eqref{eq:aupdate}):
	\begin{equation}
	P_{i,j}(\Delta R_{tot},{\bf f})=P(a_i\to \tilde{a}_i, a_j\to \tilde{a}_j)= \frac{1}{(1+  {\rm e}^{\Delta R_{tot}})(1+  {\rm e}^{\Delta s})(1+  {\rm e}^{\Delta \xi}) } \ ,
	\label{eq:prob0}
	\end{equation}
	where $\Delta R_{tot}(t)= R_{tot}(t-1)- R_{tot}(t)$  is the difference in total returns before and after the proposed move is made, $\Delta s=s(i)-s(j)$ is the difference between the security parameter of asset $i$ and $j$ and analogously $\Delta \xi= \xi(i)-\xi(j)$ is the difference between the stability parameters of the two assets. Under this functional specification, which is reminiscent of the standard Glauber dynamics in statistical physics \cite{Glauber}, there is a higher chance of transitioning to the new state if $\Delta R_{tot}(t) <0$, i.e. if there is a gain in total return by changing the weights, or if the asset's adoption chances are increased by more desirable security and stability configurations, i.e. if $\Delta s<0$, $\Delta \xi<0$.\footnote{The app may also require a fee for providing suggestions to investors and processing calculations using proprietary data stored on their servers. A  fixed {\em transaction cost} $c$ may be subtracted from the calculation of the total expected return. This cost will indeed affect $\Delta R_{tot}$ and the probability of accepting the proposed change in adoption \ref{eq:prob0} :
\begin{equation}
 \Delta R'_{tot}= R_{tot}(t-1)- R_{tot}(t) - c \ .
 \end{equation}
 If the costs associated to the proposed change exceed the potential increase in expected returns, the investor may be less inclined to make the change effective.
 Adding a transaction cost has essentially no other effects than inducing a friction term that slows down the adoption dynamics.}
  
  At each time $t$, each one of the $K$ investors proposing a change in adoption, as explained in Sec. \ref{sec:demand}, may decide to make this change effective or not based on the information calculated by the app, namely the probability in Eq. \eqref{eq:prob0}. 
	Each investor generates a uniform random threshold $p_K \in [0,1]$: the change will be made effective if $P_{i,j}>p_K$ and discarded otherwise.

	The  $K$ investors will sequentially been asked to cast their preferences (i.e. change adoption probabilities of assets $i,j$ or not). Once all $K$ investors have submitted their choices, the app will update and store the new vector ${\bf a}^{\prime}= \{a_1^{\prime}, a_2^{\prime},\dots, a_N^{\prime}\}$ of crypto assets adoption.
	
	Changes in adoption affect the expected returns of assets: at each step the app recalculates and updates the expected return for every crypto asset $i$. Here, we assume that changes in expected returns for each asset are driven by two main factors: (i) changes in the adoption rate and (ii) intrinsic features of the asset.  Specifically, we assume that the app updates the expected return of crypto asset $i$ according to the following rule: 	
	\begin{equation}
	 r_i(t) =  r_i(t-1) + \Delta a_i(t)  +\eta_i(t) \ .
	 \label{eq:returnsupdate}
	 	\end{equation}
	$ \Delta a_i(t)= a_i(t-1) - a_i(t)$ represents the change in adoption of asset time $i$ from the previous steps of the dynamics $t-1$ to the current one $t$. The term $\eta_i(t)$ represents a random component or noise generated from a Gaussian distribution with mean $\mu=0$ and variance $\sigma_i= f(\xi_i)$, which is a function of the stability parameter $\xi_i$ of asset $i$. In the following we will define  $ f(\xi_i)= \frac{1}{\xi_i}$: the higher is the {\em stability} of the asset, the smaller will the fluctuations in expected returns be. 

\subsubsection{Optimal recommendations} \label{sec:optima}

        	The dynamics over the adoption probabilities equilibrates at a time $t^\star$, when $a_i(t^\star) = a_i(t^\star-1)=a_i^{\star}, \quad \forall i=1,\dots, N$. This corresponds to having a small chance of accepting any new proposed state in probability of adoption for all the investors over all assets pairs $i,j$:      	
        	\begin{equation}
        	P_{i,j}-p_K <\epsilon \quad \forall i,j \ .
        	\end{equation}	
	
	According to Eq. \eqref{eq:returnsupdate}, even when the adoption probabilities have stabilised, i.e. at $t^\star$, the expected returns would still be randomly fluctuating, due to the random noise $\eta_i(t^{\star})$:
	\begin{equation}
	r_i^{\star}:= r_i(t^{\star}) =  r_i(t^{\star}-1) +\eta_i(t^{\star})  \ .
	 \label{eq:returnsupdatestar}
	 	\end{equation}
		
	The app will then need to provide optimal recommendation for the volatility of the expected returns. We assume that this recommendation is computed over the different $\kappa=1,\dots,4$ classes of assets we considered in this model. The way this is done is by attempting to maximise the total expected return of each sub-class while simultaneously keeping its volatility bounded by minimising the distances of each asset from the ``centre of mass" of their respective clusters. 
	
	At $t^\star$ the app computes the mean adoption probability $\bar a^{\star}= \frac{1}{N_\kappa}\sum_{i \in\kappa} a_i^{\star}$ and mean return $\bar r^{\star}=\frac{1}{N_\kappa}\sum_{i \in\kappa} r_i^{\star}$ for each class $\kappa$, and then runs the following optimisation problem per asset class to find the pair $(a_i^{\rm fin}, r_i^{\rm fin}), \forall i$:
	
	\begin{eqnarray}
	\min\limits_{a_i, r_i, i \in \kappa}  \sqrt{(a_i- \bar{a}^{\star})^2+(r_i - \bar{r}^{\star})^2} \ ,\nonumber \\
	{\rm s.t.} \sum_{i\in \kappa} r_i a_i \geq \sum_{i\in \kappa} r_i^{\star} a_i^{\star} \ .
	\label{eq:optimise}
	\end{eqnarray}
	
	All these steps can be interpreted as part of an optimal selection process of crypto assets performed by the investors using information and recommendation provided by the crypto app. At the end of the selection crypto assets will naturally cluster in different regions of the adoption-return space, depending on their features and the investors' preferences.
		
\begin{figure}[H]
\centering
\includegraphics[width=0.45\textwidth]{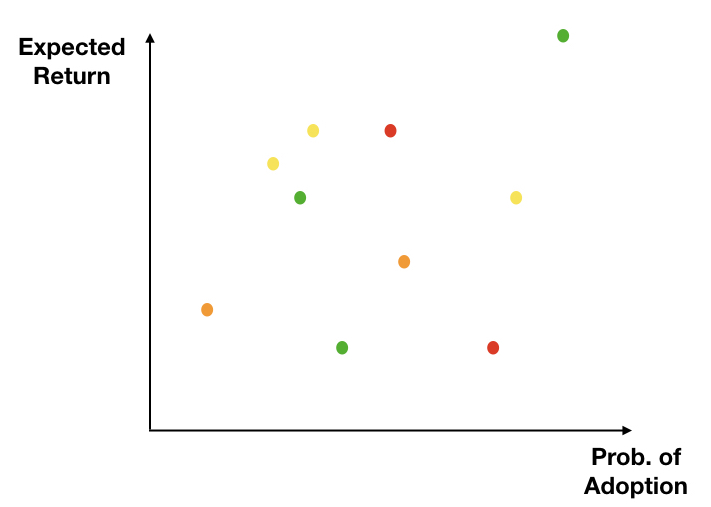}
\includegraphics[width=0.45\textwidth]{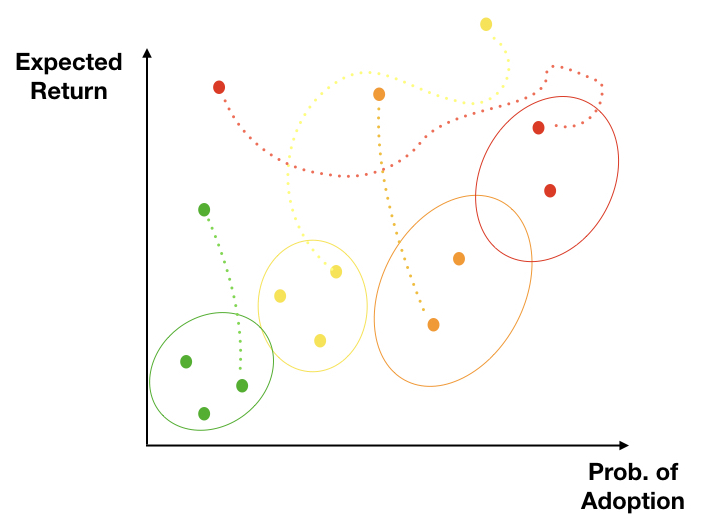}
\caption{Schematic representation of the crypto assets dynamics.}
\label{fig:cryptodyn}
\end{figure}

	A schematic representation of the crypto assets dynamics is presented in Fig. \ref{fig:cryptodyn}. 
	
In Fig. \ref{fig:cryptodyn}, we schematically describe how assets may move in the adoption-expected return space starting from the initial configuration as their values are updated. As users start interacting with the app and taking advantage of its recommendations, the local knowledge on the assets, namely their adoption probabilities $a_i(t)$ and their expected returns $r_i(t)$, as well as the global state are updated. Different colours represent assets belonging to different classes, depending on their attributes.

\subsubsection{Settings}  \label{sec:settings}
	The app can accommodate user-specific settings. Each user can define the weights it places on assets features and global variables so that the acceptance probability in Eq. \eqref{eq:prob0}  can reflect investors' preferences, for example as follows:
	\begin{equation}
	P(a_i\to \tilde{a}_i, a_j\to \tilde{a}_j)= \frac{1}{(1+  {\rm e}^{\beta_0\Delta R_{tot}})(1+  {\rm e}^{\beta_1\Delta s})(1+  {\rm e}^{\beta_2\Delta \xi}) } \ ,
	\label{eq:prob}
	\end{equation}
	where the $\beta$s parameters represent investors' attitudes towards security, stability of individual assets as well as a global state of the system. Those parameters are used to tune (e.g. increasing or decreasing) the importance of individual assets attributes.	
\section{Simulations results} \label{sec:result}
In this section, we simulate $K$ investors interacting with $N$ crypto assets via a crypto app and we analyse possible outcomes for the crypto assets market in terms of adoption and expected returns. We consider different scenarios in terms of investors' attitudes towards assets' attributes. In Sec. \ref{sec:homo}, we analyse the case of homogeneous investors, all characterised by the same parameters $\beta_i, i=0,1,2$. In the context of the crypto app setting, this means that all users decide to utilise parameters set up by the app itself, or that the app calculates average $\beta$s using the input of all users. In Sec. \ref{sec:hetero}, we extend the model to consider heterogeneous investors with parameters $\beta$s extracted from different probability distributions. In the app context, each investor sets up its own level of preferences.

	We initialise the dynamics by creating $N$ different crypto assets $\kappa_i, i=1,\dots,N$, where each asset $i$ is defined by a {\em vector of intrinsic features} ${\bf f}_i$.
	Features for each asset $s_i, \xi_i, i=1, \dots,N $ can assume a value in $[0,1]$, and will be randomly generated from  $\pi(s), \hat{\pi}(\xi)$ respectively, the probability density functions of the security and stability parameters. Features can be extracted from different distributions, yielding a different set of assets investors can buy or sell. In this way, we generate different types of assets, belonging to the four main subfamilies. 
	
In these simulations, the assets features will not evolve in time or adapt, but will be considered fixed throughout the adoption and investment process. Moreover, the number of assets $N$ available to investors will remain constant, assuming that in the timescale of observation of the market no new assets will be created nor existing ones will disappear (due to default or failure of the platform). 

The dynamics is in discrete time and simulations are run for $n_s$ steps until convergence. It is important to note that for the sake of the simulations, to guarantee convergence in the return space we will not exactly calculate the full optimisation at every step as described in Eq. \eqref{eq:optimise}, but we will implement an effective process by rescaling of an arbitrary fixed quantity the adoption and return values across different classes at each step, until the distance minimisation condition is satisfied.

	Specifically, we use the following protocol. At each step $t$:
	\begin{enumerate}
	\item we compute the``centre of mass" for $N_\kappa$ assets belonging to the same $\kappa$ class, 
	\begin{equation}
	(\bar{a}(t), \bar{r}(t))=\left (\frac{1}{N_\kappa}\sum_{i\in \kappa}  a_i(t), \frac{1}{N_\kappa}\sum_{i\in\kappa}r_i(t)\right) \ .
	\label{eq:CM}
	\end{equation}
	\item  We check whether the distance of asset $i$ from the centre of mass,  $d_i= \sqrt{(a_i(t) - \bar{a}(t))^2+(r_i(t)- \bar{r}(t))^2}$, exceeds a fixed threshold $\theta$.
	\item If $d_i \geq \theta$ we adjust both $r_i(t)$ and $a_i(t)$ as follows
	\begin{eqnarray}
	r_i^{ adj}(t) &= r_i (t) - \frac{\bar{r}(t)- r_i(t)}{2}\ , \\
	a_i^{ adj}(t) &= a_i(t) - \frac{\bar{a}(t) - a_i(t)}{2} \ .
	\end{eqnarray}
	\end{enumerate}
 We iterate this process for all assets until convergence, when $r_i^{\rm fin}=r_i^{ adj}(t^\star)$ and $a_i^{\rm fin}=a_i^{ adj}(t^\star)$ defined in Sec. \ref{sec:optima}.

\subsection{Homogeneous investors case}\label{sec:homo}
We simulate different crypto-ecosystems by assuming that the assets features -- security $s$ and stability $\xi$ -- are both generated from  uniform probability distributions $\pi(s), \hat{\pi}(\xi)$ in $[0,1]$. We consider a representative investor (all investors have the same parameters $\beta_{0,1,2}$) with different propensity levels ($\beta_1,\beta_2$) for the two types of features,  security and stability, and we observe the outcome of the agent's decisions.

\begin{figure}[htb!]
         \centering
        \includegraphics[width=0.55\textwidth]{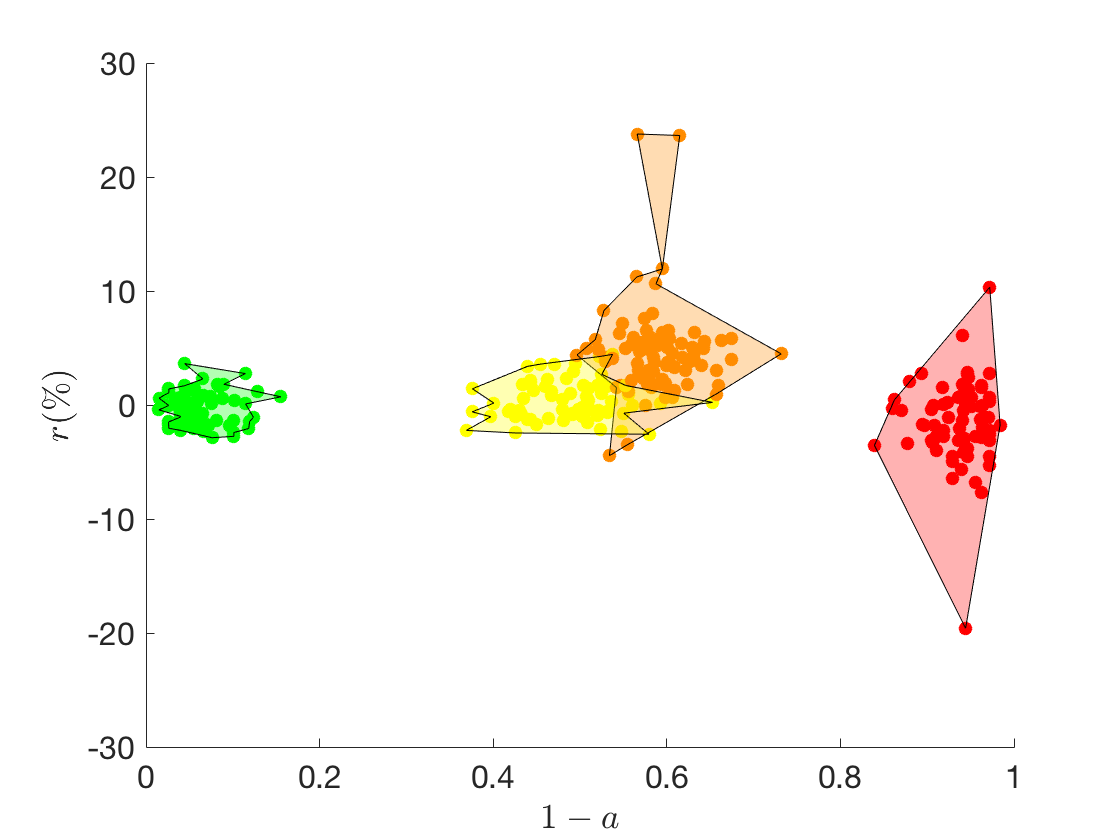}
         \includegraphics[width=0.6\textwidth]{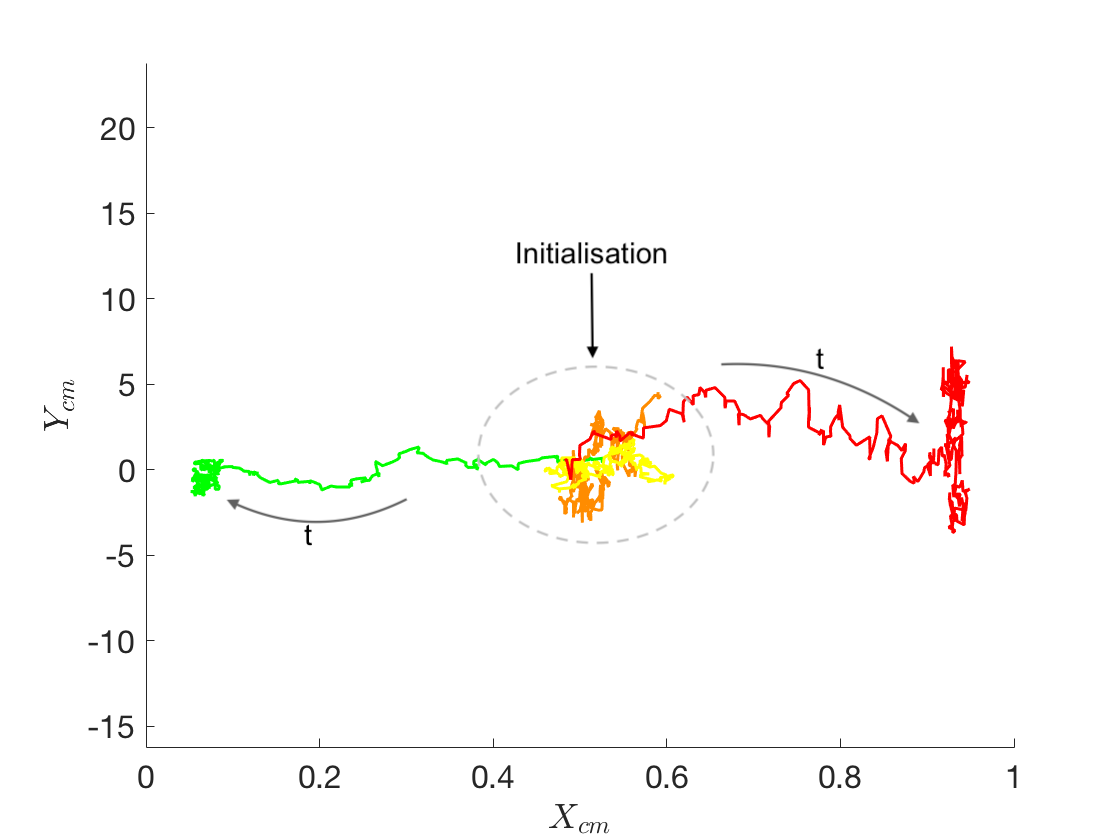}

        \caption{Top: Optimal assets selection outcome in the adoption-expected return space. Simulation with $N=300$ crypto assets, $\delta=0.1$, $\beta_0=\beta_1=\beta_2=1$ and $n_s=400$. Bottom: Evolution of centre of mass trajectories in time for the different asset classes. {\em Legend} -- green: CBDCs, orange: cryptocurrencies, red: crypto tokens, yellow: stable coins.}
        \label{fig:T1}
         \end{figure}

\begin{figure}[htb!]
         \centering
                \includegraphics[width=0.47\textwidth]{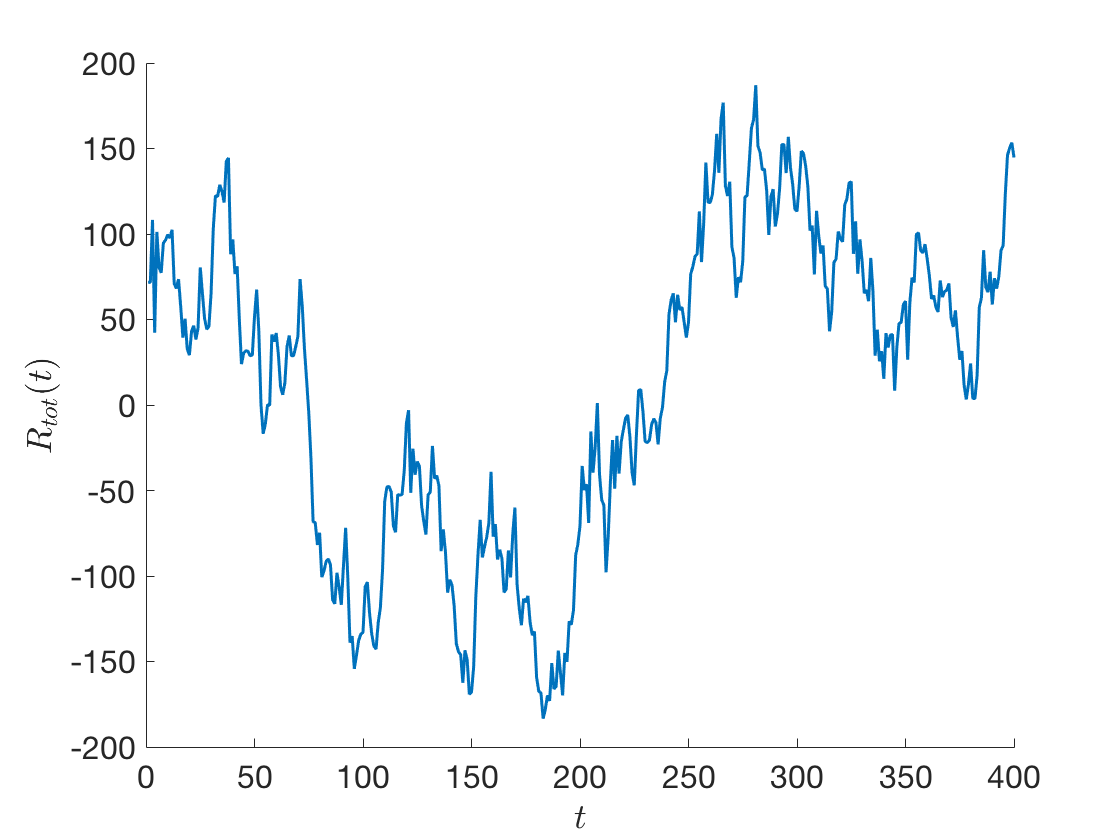}
                                \includegraphics[width=0.47\textwidth]{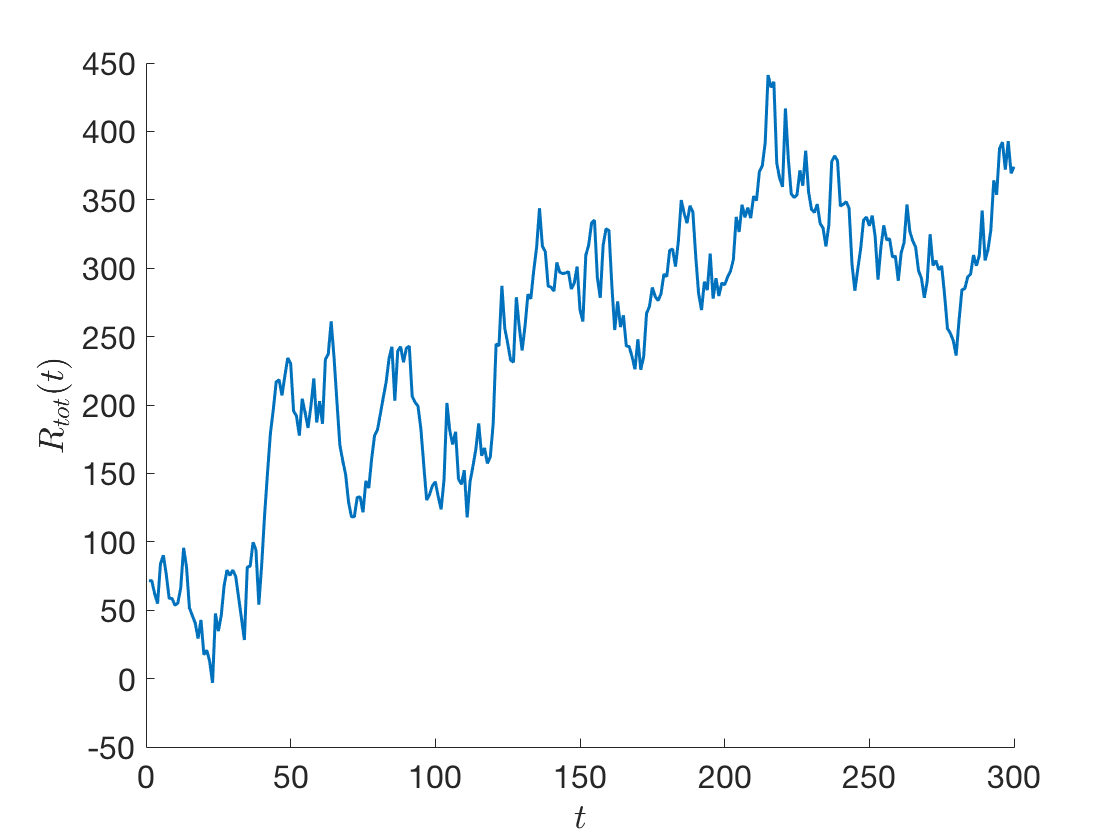}
        \caption{Total expected return in time for different $\beta$ parameters. Left: Simulation with $N=300$, $\delta=0.1$, $\beta_0=1$, $\beta_1=\beta_2=1$ and $n_s=400$. Right: Simulation with $N=300$ , $\delta=0.1$, $\beta_0=1$, $\beta_1=\beta_2=0.01$ and $n_s=300$.}
        \label{fig:CMTR1}
         \end{figure}

Let us first consider the case of a representative investor with $\beta_0=\beta_1=\beta_2=1$. In this scenario, all factors -- total expected return, security and stability parameters of the assets --  contribute equally to the acceptance probability in Eq. \eqref{eq:prob}. The representative investor will, indeed, change the assets' adoption probabilities (according to Eq. (\ref{eq:aupdate})) by optimising both with respect to the total return and the assets' parameters, security and stability. Equivalently, the change will be accepted if there is an increase in total return, and if the asset whose adoption is increased is more secure and stable than the one we are comparing it with. Under these conditions, as shown in Fig. \ref{fig:T1}, the assets most likely to be adopted in the future are CBDCs and stablecoins, while cryptocurrencies and crypto tokens are the least adopted but also the ones with the highest fluctuations in expected return (less stable).  Indeed, given that $\beta_1=\beta_2=1$, there is neither a strong aversion nor propensity from the investor's point of view towards security or stability features of the assets.
      In Fig. \ref{fig:T1} (lower panel), we monitor the convergence via the dynamics of the centre of mass (left), as defined in Eq. \eqref{eq:CM}. In Fig. \ref{fig:CMTR1} (left) we show the evolution in time of the total expected return (right) (see Eq. \ref{eq:rtot}). 
      
\begin{figure}[htb!]
         \centering
        \includegraphics[width=0.6\textwidth]{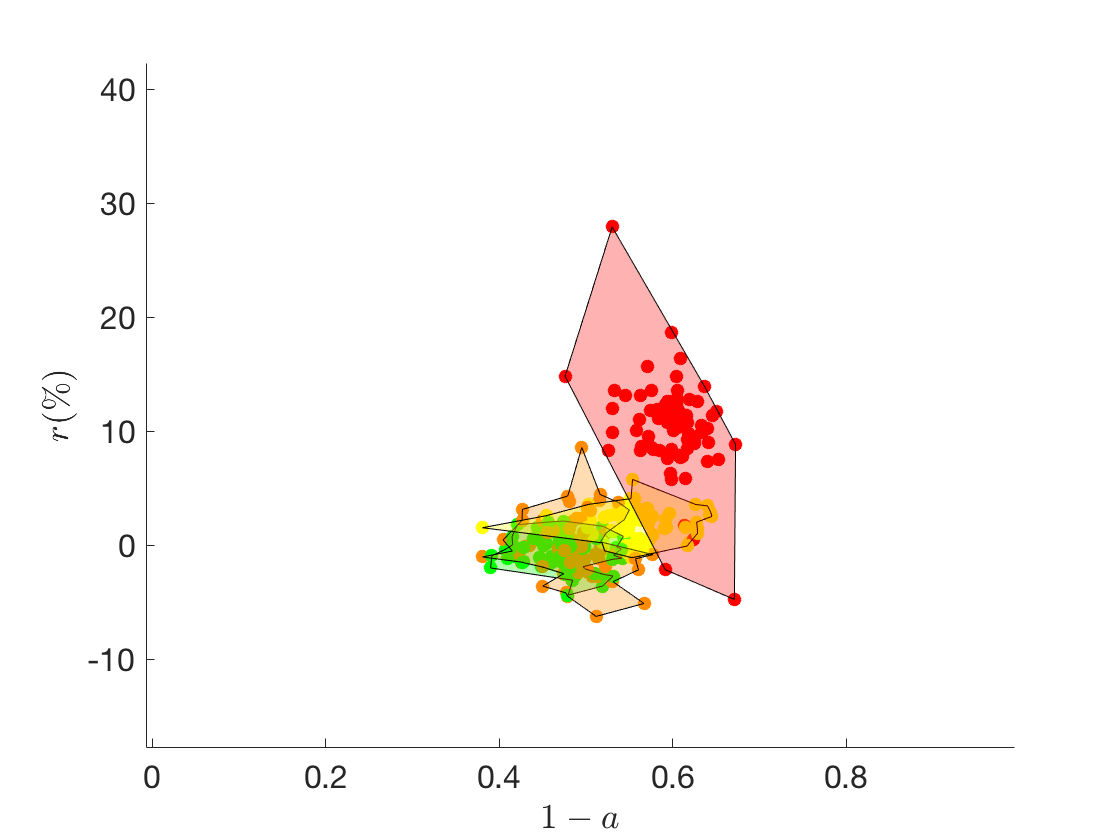}
        \caption{Optimal assets selection outcome in the adoption-expected return space. Simulation with $N=300$ crypto assets, $\delta=0.1$, $\beta_0=1$, $\beta_1=\beta_2=0.01$ and $n_s=300$. {\em Legend} -- green: CBDCs, orange: cryptocurrencies, red: crypto tokens, yellow: stable coins.}
        \label{fig:T2}
         \end{figure}

   By choosing a very small  parameter $\beta$, the dynamics is essentially driven by the maximisation of the total return and will not be strongly affected  by the intrinsic assets' features. In Fig. \ref{fig:T2}, we show the assets dynamics for $\beta_1=\beta_2=0.01$: all assets mean adoption probabilities are centred around $a_i \sim 0.5, \forall i=1,\dots,N$ (i.e. they will have a $50\%$ chance of surviving) indicating that the representative investor is not biased towards investing or not on a given asset $i$ (based on the asset's features). The dynamics is mostly driven by fluctuations in total return, as is also noticeable from the average behaviour of each asset class in figure \ref{fig:CMTR1} (right).    
   By observing the total return in time $R_{tot}(t)$ in Fig. \ref{fig:CMTR1} (right), we clearly notice, despite the stochasticity, an upward trend in time, differently from the case with $\beta>0$, where the optimisation over the assets' features as well led to a less steady increase of the total return for the representative investor (see Fig. \ref{fig:CMTR1}, left panel). As shown in figure \ref{fig:T2}, the assets will differentiate mostly in terms of their expected return, with cryptocurrencies and crypto tokens experiencing the widest fluctuations (due to their lower stability parameters $\xi$s).
         
      Note that the results are not an artefact of the choice of the acceptance probability in Eq. \eqref{eq:prob0}: different functional forms with similar monotonic behaviour yield qualitatively similar results. In our model, we opted for a transition probability in a form compatible with a Glauber dynamics \cite{Glauber}: we essentially set up a Markov Chain, or equivalently a local dynamical rule specifying under which conditions a system in an initial state $\mu$ should  transition to a new state $\nu$. By making a parallel with statistical physics, assets in our space are moving on a rough landscape and they are exploring it, via the Glauber dynamics, searching for global minima of their complex energy function. In some cases a move that decreases the total returns (i.e. $\Delta R_{tot}(t) > 0$), or which increases the adoption probabilities of riskier assets (e.g. $\Delta s>0$) might still be accepted with a certain probability, according to Eq. \eqref{eq:prob0}: indeed, it is often necessary to move against the gradient that would push the asset towards a local minimum, to fully explore the entire landscape and discover a deeper energy valley. Consistently with the parallel with statistical physics, the $\beta$-parameters in Eq. \eqref{eq:prob} act as an equivalent of an inverse temperature \cite{huang}. 
      
\begin{figure}[htb!]
         \centering
        \includegraphics[width=0.6\textwidth]{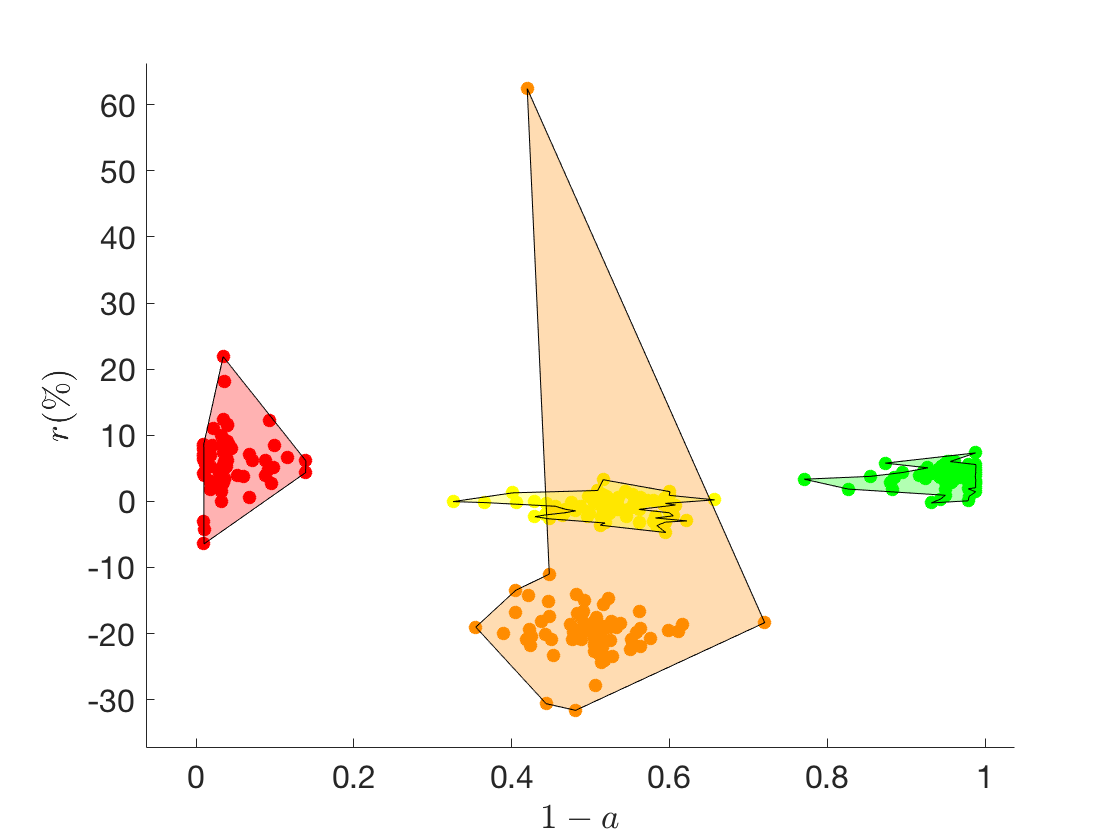}
        \caption{Optimal assets selection outcome in the adoption-expected return space. Simulation with $N=300$ crypto assets, $\delta=0.1$, $\beta_0=1$, $\beta_1=\beta_2=-2$ and $n_s=500$. {\em Legend} -- green: CBDCs, orange: cryptocurrencies, red: crypto tokens, yellow: stable coins.}
        \label{fig:T3}
         \end{figure}

  The parameter $\beta$ essentially defines the investor(s)'s attitudes. In particular, by setting $\beta < 0$ we model a risk prone investor, who will be more likely to invest on less secure and/or less stable assets. For instance, in figure \ref{fig:T3}, we analyse the case of $\beta_1=\beta_2=-2$, where the investor has a strong propensity towards both types of features. In this case, the stable configuration for the assets is characterised by the riskiest assets, i.e. cryptocurrencies and crypto tokens, having a low probability of not being adopted. The outcome is clearly specular to the one observed for a risk averse investor in Fig. \ref{fig:T1}.
   
         \begin{figure}[htb!]
         \centering
        \includegraphics[width=0.47\textwidth]{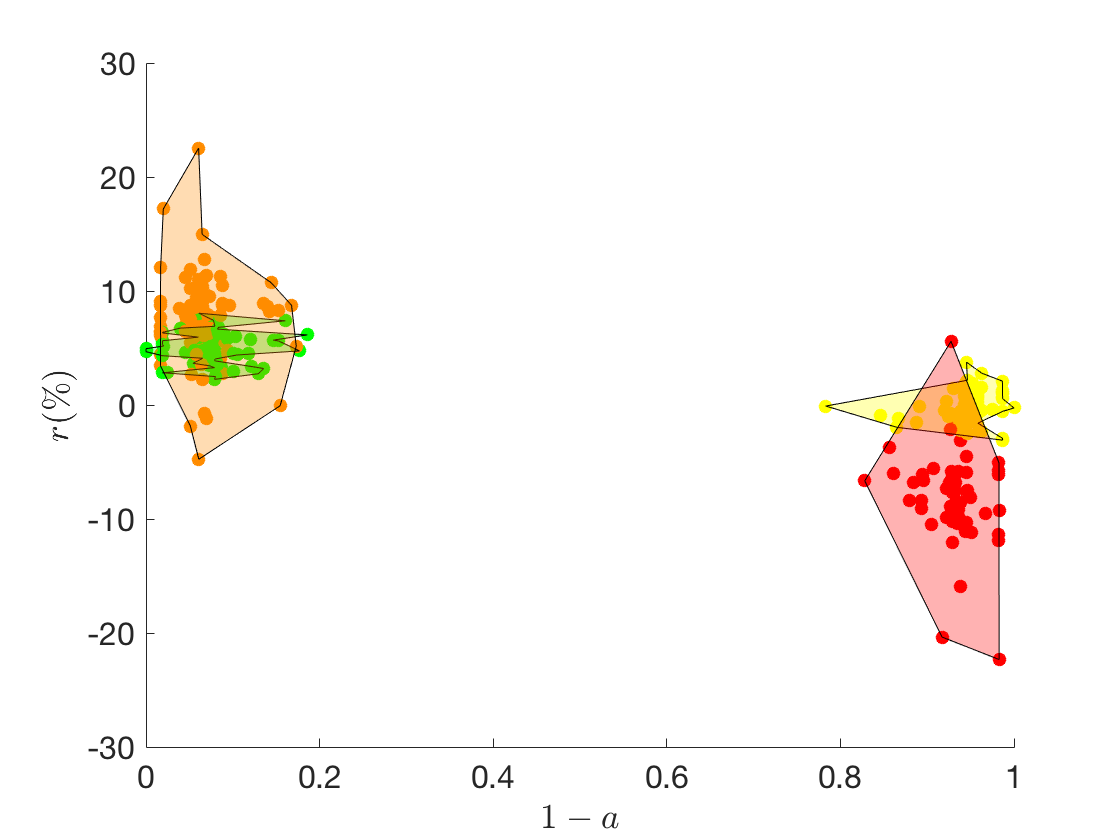}
         \includegraphics[width=0.47\textwidth]{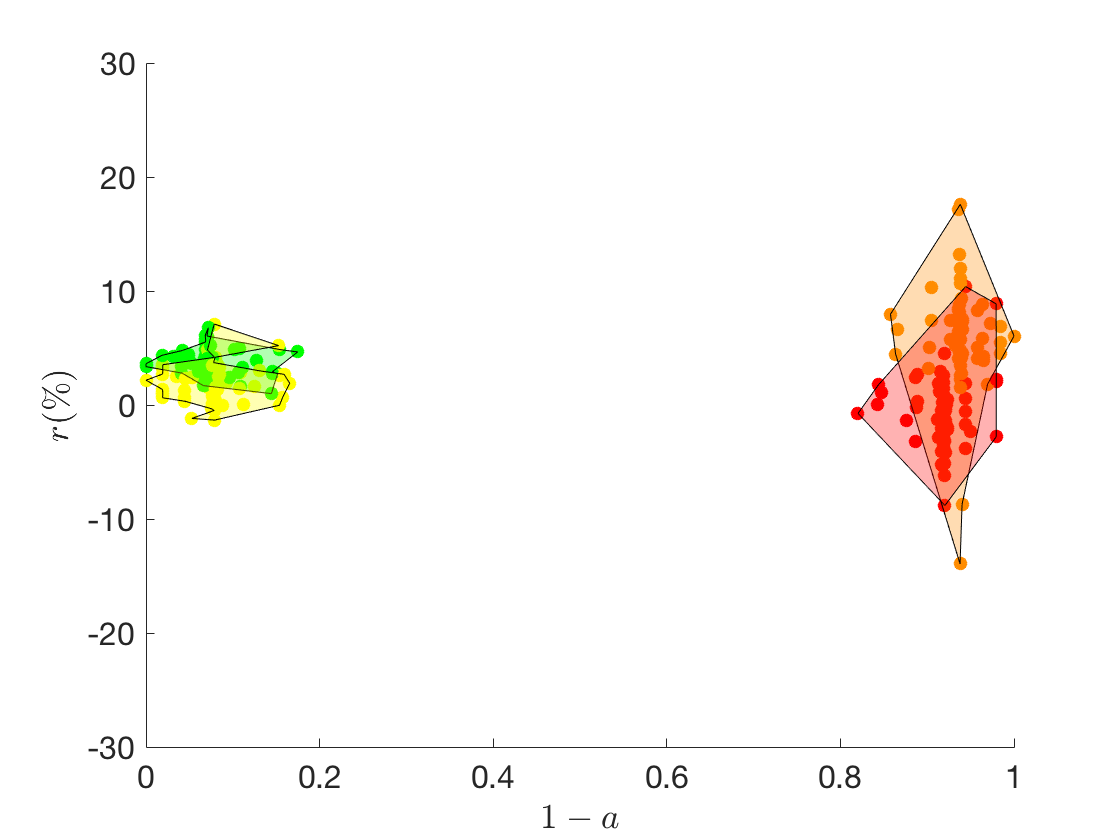}
        \caption{Optimal assets selection outcome in the adoption-expected return space. Simulation with $N=300$ crypto assets, $\delta=0.1$ and $n_s=300$, $\beta_0=1$. Left: $\beta_1=2, \beta_2=0.01$. Right: $\beta_1=0.01, \beta_2=2$. {\em Legend} -- green: CBDCs, orange: cryptocurrencies, red: crypto tokens, yellow: stable coins.}
        \label{fig:betadiff}
         \end{figure}
       Our representative investor on the crypto app may also have different attitudes towards the different assets' features. For example, an investor may be more (or less) prone to tolerate risks associated to the assets' security or stability attributes: this would correspond to setting $\beta_1\neq\beta_2$ in the crypto app user's preferences. Indeed, according to Eq. (\ref{eq:prob}), $\beta_1$ will represent the risk aversion parameters towards risks associated to the asset's {\em security} features, while $\beta_2$ will take into account risks connected to the asset's {\em stability}.

        In Fig. \ref{fig:betadiff}, we compare two opposite cases: one where the representative investor has a high aversion towards to low-stability assets(high $\beta_1$), while considerations on the security of the assets do not affect the probability that a given investor will adopt them (low $\beta_2$, close to zero) with the mirroring one, with a high $\beta_2$ and a negligible $\beta_1$. In the first case (Fig. \ref{fig:betadiff}, left panel), the most adopted assets are indeed also the most secure ones according to our classification, i.e. CBDCs and cryptocurrencies. In the second case (Fig. \ref{fig:betadiff}, right panel), the representative investor has instead a predilection for more stable assets, which correspond to stablecoins and CBDCs and which now occupy the leftmost part of the diagram.

\begin{figure}
         \centering
        \includegraphics[width=1.1\textwidth]{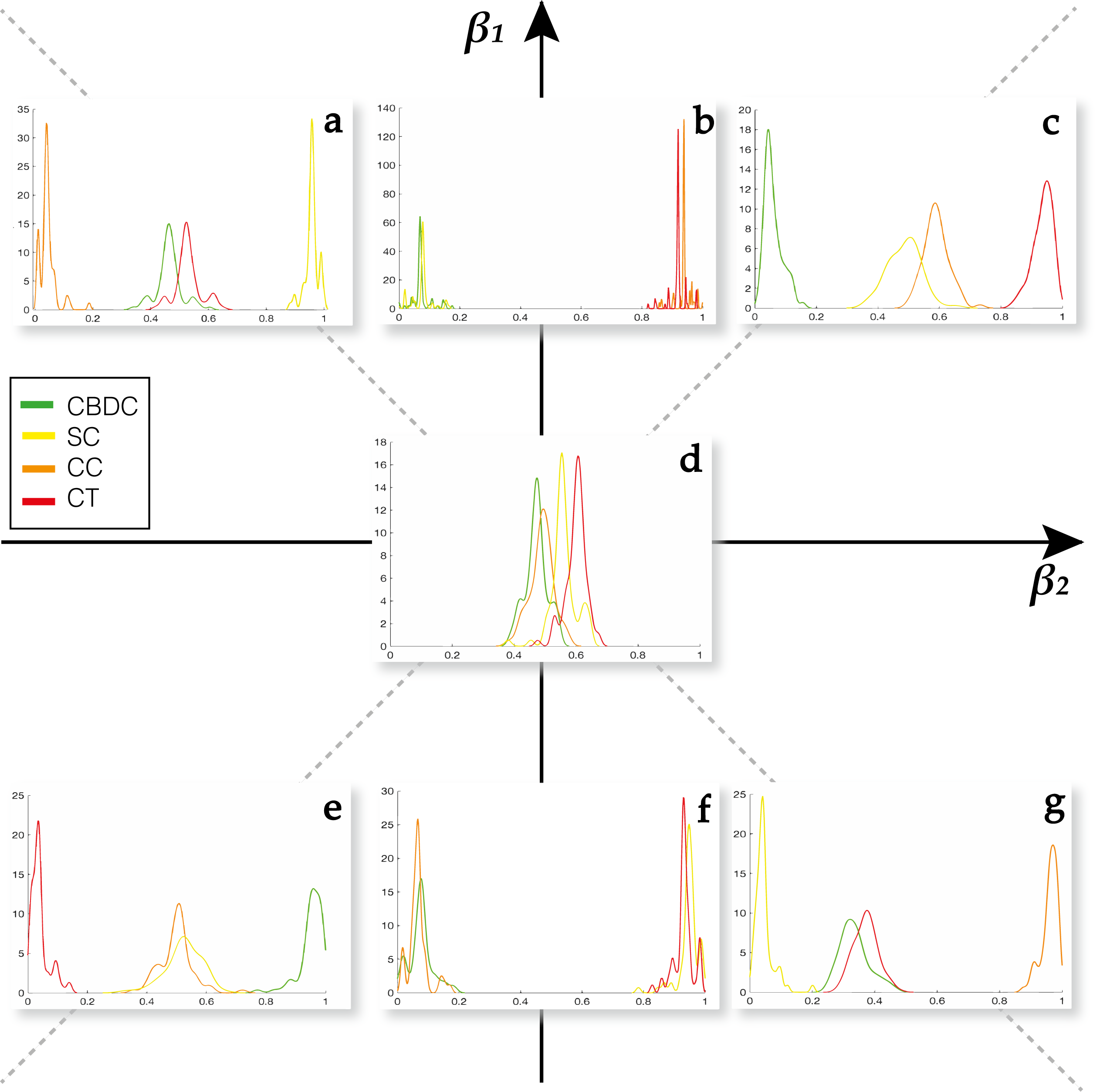}
        \caption{Phase diagram of the system varying the parameters $\beta_1, \beta_2$ with $\beta_0=1$. By monitoring the distribution of mean probability of not being adopted $1-a$, conditioned on the assets' class $\kappa$, $P(1-a |\kappa)$, we can identify different stable configurations for the system, in different regions of the parameter space. We include few illustrative examples: a) $\beta_1= -2 , \beta_2=2$ b) $\beta_1= 2, \beta_2=0.01$  c) $\beta_1= \beta_2 = 2$  d) $\beta_1=  \beta_2=0.01$  e) $\beta_1= \beta_2=-2$ f) $\beta_1=-2, \beta_2=0.01 $ g) $\beta_1= 2 , \beta_2=-2$. In all simulations we consider $\delta=0.1$, $n_s=300$, $N=300$ and $\beta_0=1$.}
        \label{fig:PD}
         \end{figure}

       As we can conclude from these examples, the assets' dynamics displays a very rich behaviour by varying the system's parameters. In particular, the dynamics of the adoptions appears to be quite informative when predicting the chance of an asset of surviving in the future. 
       To better monitor changes in adoption for the different assets under various conditions, we provide a normalised histogram of $1- a $ (i.e. the mean probability of not being adopted), conditioned on the asset class (i.e. CBDCs, CTs, SCs, CCs). The results are summarised in a phase diagram-like representation as a function of the parameters $\beta_1,\beta_2$ in Fig. \ref{fig:PD}. 
       By varying the $\beta$ parameters, the ecosystem moves from one phase to another, characterised by different outcomes for the probability of not being adopted per asset class. 
       
       \subsubsection{Distribution of the assets' features and rescaling}
       So far, assets features were generated from uniform probability density functions $\pi(s), \hat{\pi}(\xi)$, so that all sub-classes were homogeneous in terms of number of assets per class. Indeed, different distributions of assets features yield a different composition of the crypto-ecosystem for the investors and this may affect the outcomes of the investment decisions. Interestingly, we can show that changes in the composition of the crypto-market can be rebalanced by modifying the investors' $\beta$ parameters in such a way that the same outcome is obtained in terms of adoption probability-expected return of the assets. 
   	Comparing two ecosystems where features were generated from $\pi(s), \hat{\pi}(\xi)$ and $\pi^{\prime}(s), \hat{\pi}^{\prime}(\xi)$, one sees that
	the acceptance probability in Eq. \eqref{eq:prob}, hence the possible outcomes for the adoption probabilities of the different asset classes may change.
	For simplicity, we consider $\beta_1=\beta_2$ in Eq. \eqref{eq:prob} and we ignore the returns distribution in the acceptance probability for the following estimation: $P\propto \frac{1}{(1+  {\rm e}^{\beta\Delta s})(1+  {\rm e}^{\beta\Delta \xi})}$.
	Given $\beta$ in the first ecosystem, what would be $\beta^{\prime}$ for the second ecosystem, yielding the same acceptance probabilities, i.e. $P= P^{\prime}$? We can produce a rough estimation of $\beta^{\prime}$ by imposing: 
	\begin{equation}
	\left\langle\left((1+  {\rm e}^{\beta\Delta s}\right)\left(1+  {\rm e}^{\beta\Delta \xi}\right)\right\rangle_{\pi(s), \hat{\pi}(\xi)} \stackrel{?}{=} \left\langle\left(1+  {\rm e}^{\beta^{\prime}\Delta s}\right)\left(1+  {\rm e}^{\beta^{\prime}\Delta \xi}\right)\right\rangle_{\pi^{\prime}(s), \hat{\pi}^{\prime}(\xi)} \ ,
	\label{eq:betaest}
	\end{equation}
	where we average over the probability density functions of the two features in the two ecosystems. The average can be explicitly written as follows
\begin{eqnarray}
&\left\langle\left(1+  {\rm e}^{\beta\Delta s}\right)\left(1+  {\rm e}^{\beta\Delta \xi}\right)\right\rangle_{\pi(s), \hat{\pi}(\xi)}\simeq \\
&= \sum_{j,k} \frac{n_j}{N} \frac{n_j-\delta_{i,j}}{N} \int_{S(j)}{\rm d}\pi(s)\int_{\Xi(j)}{\rm d}\hat{\pi}(\xi)\int_{T(k)}{\rm d}\pi(\tau)
\int_{R(k)}{\rm d}\hat{\pi}(\rho)(1+  {\rm e}^{\beta(s-\tau})(1+  {\rm e}^{\beta(\xi-\rho)}) \ ,& \nonumber
\label{eq:approx}
\end{eqnarray}
where $n_j= N \int_{S(j)}{\rm d} s \pi(s)\int_{\Xi(j)}{\rm d}\xi \hat{\pi}(\xi)$ is the average number of crypto assets in class $j=1,\dots,4$ and $n_j/N$ is the probability of extracting an asset of class $j$. We also introduced the following notation where ${\rm d}s\pi(s)={\rm d}\pi(s)$. We integrate over four regions $S(j),\Xi(j), T(k), R(k)$ where the stability and security parameters of the assets $j$ or $k$ are defined. For instance, extracting asset $j$ from the CBDC class would mean--as defined in table \ref{fig:tablec} --that the security and stability parameters lie respectively in the regions $s \in [0.5,1] = S(j)$ and $\xi \in [0.5,1] = \Xi(j)$. In Eq. (\ref{eq:approx}) we are approximating the average over the pdfs by summing over all possible combinations of assets $j,k$ extracted from the different classes, weighted by the probabilities that the two selected assets belong to the a given class $\frac{n_j}{N} \frac{n_j-\delta_{i,j}}{N}$.

\begin{figure}[H]
         \centering
        \includegraphics[width=0.55\textwidth]{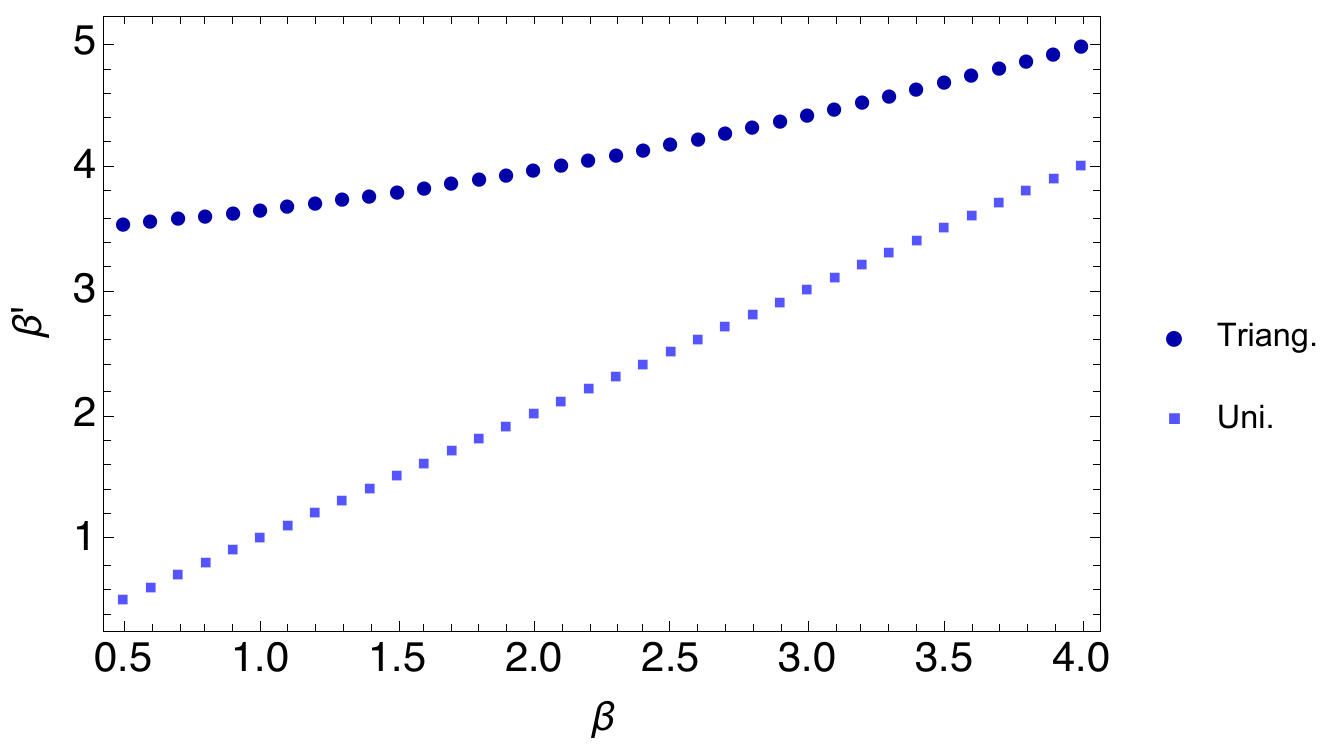}
        \caption{Numerical estimation of $\beta^{\prime}$ as a function of $\beta$ according to Eq. \eqref{eq:betaest},  given $\pi^{\prime}(s), \hat{\pi}^{\prime}(\xi)$ uniform and (i) $\pi(s)=1, \hat{\pi}(\xi)=1$ (uniform pdf with $s,\xi \in [0,1]$) and (ii)  $\pi(s)=2s, \hat{\pi}(\xi)=2\xi$ (triangular pdf with $s,\xi \in [0,1]$).}
        \label{fig:TDmathe}
         \end{figure}

\begin{figure}[H]
         \centering
        \includegraphics[width=0.6\textwidth]{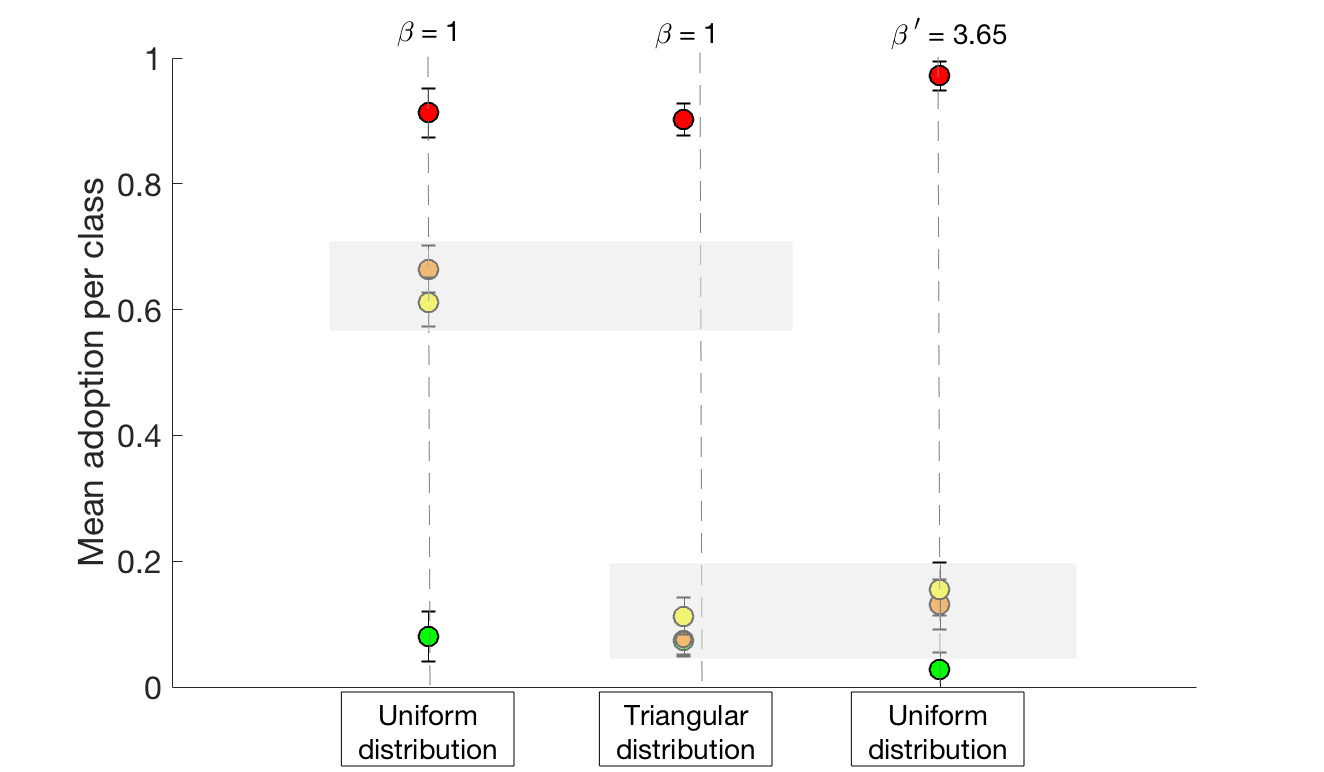}
        \caption{Mean (and variance) of the probability of not being adopted per asset class for the following scenario with $N=200$ crypto assets,  $\delta=0.1$ and (i) $\beta=1$, $\pi(s)=1, \hat{\pi}(\xi)=1$ (uniform), (ii) $\beta=1$,  $\pi(s)=2s, \hat{\pi}(\xi)=2\xi$ (triangular distribution) and (iii) $\beta^{\prime}=3.65$, $\pi(s)=1, \hat{\pi}(\xi)=1$ (uniform rescaled). }
        \label{fig:comparison}
         \end{figure}

	By solving it numerically, fixing $\beta$ we can estimate the values of $\beta^{\prime}$ that would yield the same acceptance probability, hence a similar outcome for the simulations. In Fig. \ref{fig:TDmathe}, we show the results of the numerical estimation of  Eq.\eqref{eq:betaest}. We consider the trivial case where in both ecosystems the assets' features are extracted from uniform distributions, where we recover $\beta=\beta^{\prime}$ and the case where we choose $\pi(s)=1, \hat{\pi}(\xi)=1$ in one ecosystem and $\pi^{\prime}(s)=2s, \hat{\pi}^{\prime}(\xi)=2\xi$ (triangular). This second scenario will be used in our illustrative example below and the estimated values for $\beta^{\prime}$ will be  fed into the simulations to observe the behaviour of the different ecosystems.
	Therefore, we simulate a system with $\pi(s)= \hat{\pi}(\xi)=1$ uniform between $[0,1]$ and one where the features are extracted from a triangular pdf in $[0,1]$: $\pi(s)= 2 s, \hat{\pi}(\xi)=2\xi$. In Fig. \ref{fig:comparison}, we summarise the result by plotting mean (and variance) of the probability of not being adopted per asset class. In general,  simulations with uniform and triangular distribution yield, for the same $\beta=1$, quite different outcomes. Nonetheless, by using the numerical estimation in \ref{fig:TDmathe} we find $\beta^{\prime}$ such that the outcome of the ecosystem with uniform features would become equivalent to the one generated considering a triangular pdf for the features instead.
                  
         Changes in the composition of the crypto-ecosystem, which may yield new investments and adoption outcomes, can be offset by a correct investors' assessment of the current situation and re-estimation of their attitudes towards assets' characteristics.

\subsection{Heterogeneous investors case} \label{sec:hetero}
In this section, we consider a set of $K$ heterogeneous investors with different $\beta$ parameters. At each round of the dynamics each investor picks at random a pair of assets $i,j$ and decides whether to update the adoption probabilities (see Eq. \ref{eq:aupdate}) according to their own attitudes towards assets' features.
 	Hence, we introduce two vectors $\vec{\beta}_{1,2} =  (\beta^{(1)}_{1,2},\dots,\beta^{(K)}_{1,2})$, containing the values of the investor-specific $\beta$ parameters $\beta^{(k)}_{1,2}$ for each investor $k=1,\dots,K$ using the crypto app. In the simulations, the parameters can be extracted from a probability distribution depending on the heterogeneity of the investors population that we aim to obtain. We will indicate the probability distributions, which $\vec{\beta}_{1,2}$ are extracted from, as $\hat{\Phi}_{1,2}$.
 
 In this case, the acceptance probability in Eq.(\ref{eq:prob}) can be rewritten as follows:
 \begin{equation}
	P^{(k)}(a_i\to \tilde{a}_i, a_j\to \tilde{a}_j)= \frac{1}{(1+  {\rm e}^{\Delta R_{tot}})(1+  {\rm e}^{{\beta}^{(k)}_{1}\Delta s})(1+  {\rm e}^{{\beta}^{(k)}_{2}\Delta \xi}) } , \quad k=1,\dots,K \ ,
	\label{eq:accepthetero}
	\end{equation}
	and it will depend on the investor's idiosyncratic preferences embedded in the parameters $\beta_{1,2}^{(k)}, k=1, \dots, K$. Note that in this case the acceptance probability (Eq. \eqref{eq:accepthetero}) changes from investor to investor. Interestingly, we can show that the choices of the different investors cannot be aggregated and lead to different macroscopic emergent behaviours. 
	As an illustrative example, we consider a scenario in which the values of both $\vec{\beta}_1,\vec{\beta}_2$ are drawn from a triangular distribution with support between $[-4,4]$, namely of the form $\hat{\Phi}_{1,2}(x)= \frac{x+4}{32}$. To gain an understanding of the effects of heterogeneity in investors' attitudes, we compare this scenario with the one obtained by considering only one representative investor with $\tilde{\beta}_{1,2} =\langle \beta_{1,2} \rangle = \frac{1}{K}\sum_{i=1}^K \beta_{1,2}^{(i)}$. 
		
\begin{figure}[htb!]
 \centering
 \includegraphics[width=0.45\textwidth]{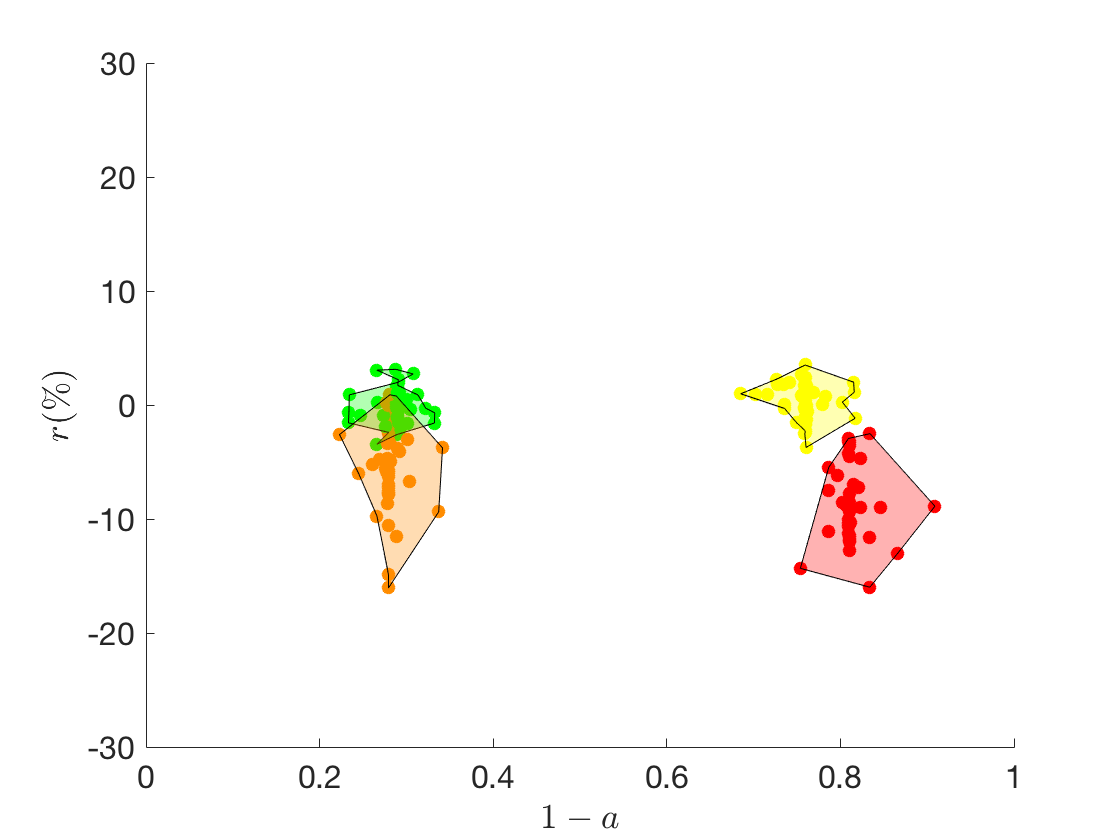}
  \includegraphics[width=0.45\textwidth]{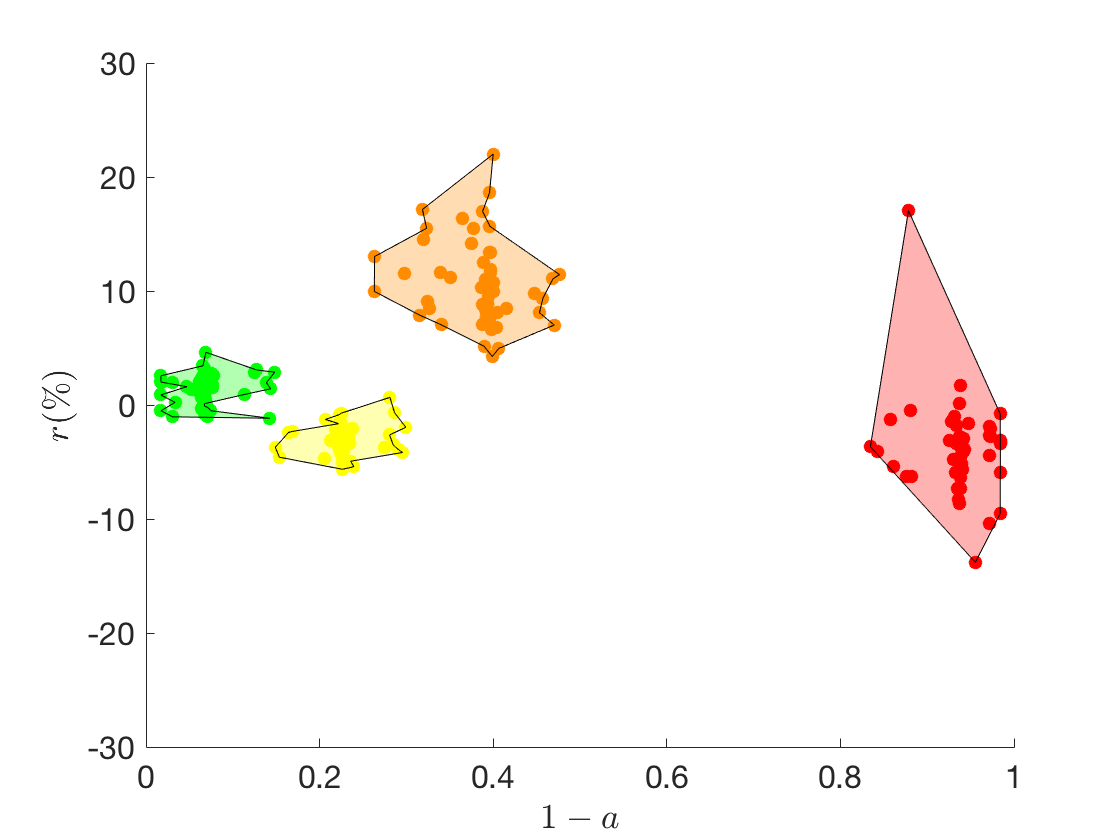}
 \includegraphics[width=0.45\textwidth]{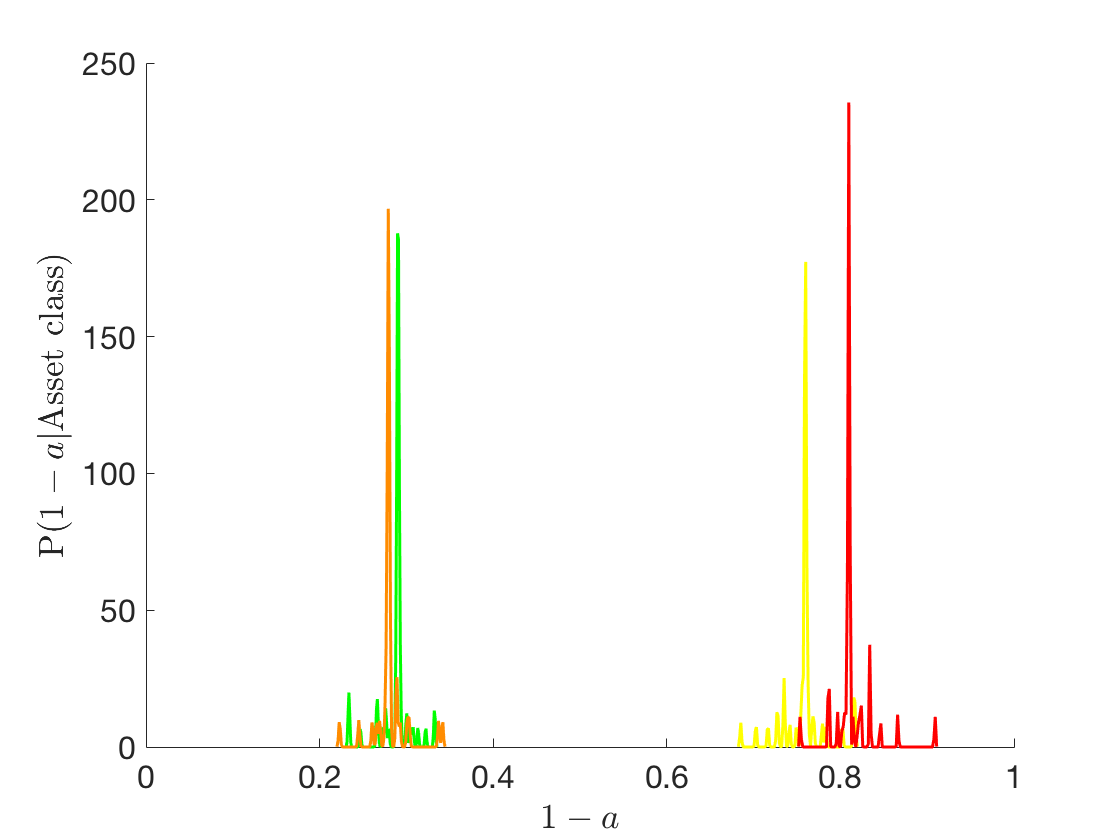}
 \includegraphics[width=0.45\textwidth]{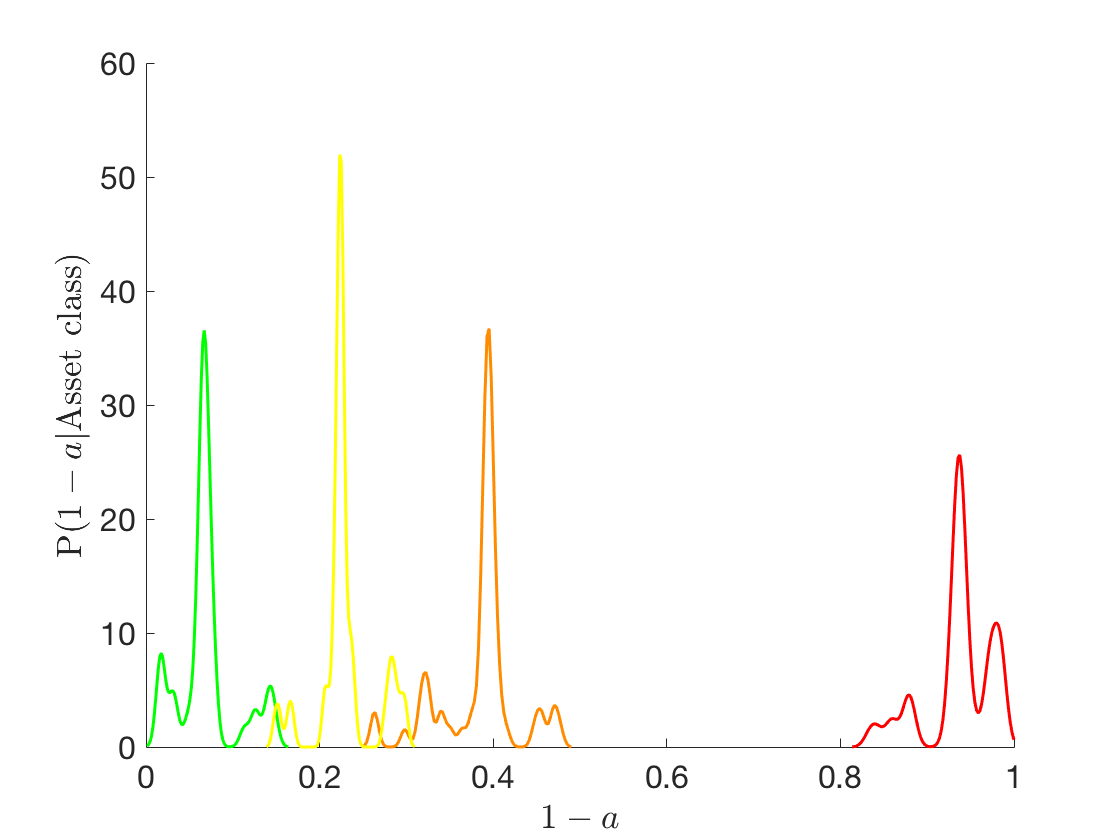}
 \caption{Simulation with $N=200$ crypto assets, $\delta=0.1$ and $n_s=300$. Top: Optimal assets selection outcome in the adoption-expected return space for the case of heterogeneous investors (left) and homogeneous investors with $\langle \beta_1\rangle= 1.234$ and $\langle \beta_2\rangle= 1.549$ (right). Bottom: Distribution of mean probability of not being adopted $1-a$, conditioned on the assets' class $\chi$, $P(1-a |\chi)$ for the heterogeneous (left) and homogeneous (right) investor case respectively.}
 \label{fig:hetvshomo}
 \end{figure}

	Results from the two simulations are shown and compared in Fig. \ref{fig:hetvshomo}. Indeed, the behaviour of the representative investors appear to be substantially different from the one observed as an outcome of the investment strategies played by misaligned investors. While in the homogeneous example the representative investor is assumed to be risk averse with respect to both types of threats (i.e. related to both security and stability) as $\tilde{\beta}_{1,2} >0$, in the heterogeneous case the investors sets is composed of both risk-prone and risk-averse agents. Due to the randomness in the process, the effects do not average out trivially, and the case of heterogeneous investors yield potentially unanticipated outcomes. For example, in the heterogeneous case, less stable or secure assets, such as crypto tokens, have a higher chance of being adopted (i.e. $a\sim0.8$) compared to the homogeneous case (with $a\sim 1$) because of the presence of more risk prone agents involved in the process (see Fig. \ref{fig:hetvshomo}).
	
In summary, by considering that all investors are characterised by the same parameters $\beta_1$, $\beta_2$, we are assuming that agents in the market have aligned strategies and similar propensity towards risk. In this ``homogeneous investors" case, the behaviour of multiple agents can be, therefore, described by a {\em representative investor} with parameters $\beta_1,\beta_2$. Introducing misaligned investors with opposite strategies determines a non-trivial behaviour in the system together with the emergence of new stable configurations in the market. Indeed, disseminating contrasting views, which may affect investors’ opinions on the assets and the associated risks, may destabilise the ecosystem yielding unexpected scenarios for the investments.	

\section{Concluding remarks} \label{sec:conclusions}

We offer a framework to classify crypto assets into major stable sub-classes in terms of two main intrinsic asset's features, namely the asset's security and stability. By considering the behaviour of different types of investors, driven by their attitudes towards assets' attributes ($\beta$ parameters), we explore different outcomes for the investments in the crypto-ecosystem and the future--in terms of their adoption probability, hence their chances of surviving-- of the crypto assets.

For the time being, we fixed the number of assets, the number of investors, the assets' features, and investor preferences throughout the analysis. In principle, any of these features could dynamically evolve together with the asset's adoption probabilities and expected returns. For instance, an asset initially characterised by low security but experiencing an increase in adoption in time, may gradually try to improve its intrinsic features (e.g. investing on improving the underlying technology) to attract even more interest.  Also, the number of assets may vary in time and this feature may be introduced in a further extension to the model. Assets with features that would fit the most the investments trends and market demand will survive the analogue of a natural selection process for the crypto ecosystem.

Our work paves the way for further investigations whose scope is to understand the future developments of the fast-growing crypto asset ecosystem. This initial exploration lays the foundation for designing successful investment strategies and predicting adoption scenarios under different conditions.

\end{document}